\documentclass[12pt]{article}

\makeatletter
\def\alt{\mathrel{\mathpalette\gl@align<}}
\def\agt{\mathrel{\mathpalette\gl@align>}}
\def\gl@align#1#2{\lower.6ex\vbox{\baselineskip\z@skip\lineskip\z@
\ialign{$\m@th#1\hfil##\hfil$\crcr#2\crcr\sim\crcr}}}
\makeatother

\begin{document}

\baselineskip 18pt

%
\vspace*{1.0cm}
\begin{center}
\baselineskip 20pt
{\Large\bf
Nearly right unitarity triangle and CP phase\\
in quark and lepton flavor mixings
} \vspace{1cm}

{\large
Yukihiro Mimura}
\vspace{.5cm}

{\it
Department of Physics, National Taiwan University, Taipei 10617, Taiwan
}
\\

\vspace{.5cm}

\end{center}

\begin{center}{\bf Abstract}
\end{center}

The nearly right unitarity triangle can be simply obtained
if a quark mass matrix is written as a linear combination of real rank-1 matrices
and the coefficient which gives the mass of the second generation
is pure imaginary.
Supposing that the source of the CP violation is in the Yukawa coupling
to an additional Higgs field which provides a factor for the strange quark and muon mass ratio,
we obtain that the angle $\alpha= \phi_2$ of the unitarity triangle shifts from $90^{\rm o}$ 
by units of $V_{ub}/V_{us}$, and 
is postdicted as $\alpha \simeq 87^{\rm o}$ or $91^{\rm o}$.
The Dirac CP phase $\delta$ which appears in the three-flavor neutrino oscillations is 
obtained to be $|\delta| \simeq 80^{\rm o}$
if the neutrino mass matrix gives tri-bimaximal-like mixing form.
If the factor 3 for the muon and strange quark mass is considered in a simple manner
in the quark-lepton unification,
we obtain three distinctive prediction of $\delta$ for nearly right-angled phase as
$|\delta| \simeq 70^{\rm o}$, $90^{\rm o}$, or $110^{\rm o}$
in an idealistic orthogonal structure of the neutrino mass matrix.
The deviation from $90^{\rm o}$ is roughly given by $\arcsin(1/3)$
by inputting the experimental measurements.

\thispagestyle{empty}

\bigskip
\newpage

\addtocounter{page}{-1}

\section{Introduction}

CP violation in the flavor mixings is one of the important topic in particle physics.
Not only in the quark sector, but also in the lepton sector,
the CP phase is being measured accurately.
The unitarity triangle of Cabibbo-Kobayashi-Maskawa (CKM) quark mixing matrix
seems to be nearly right-angled.
The measurement of one of the angles is 
\cite{Patrignani:2016xqp}
\begin{equation}
\alpha\ (=\phi_2)= (87.6 ^{+3.5}_{-3.3})^{\rm o}.
\end{equation}
The triangle in 
Pontecorvo-Maki-Nakagawa-Sakata (PMNS) neutrino mixing matrix
also seems to be nearly right-angled.
Conventionally, Dirac CP phase $\delta$, which appears in the neutrino oscillations,
is concerned in the lepton sector, and 
the recent global data fits of the neutrino oscillations give
\cite{Capozzi:2018ubv,Esteban:2016qun}
\begin{equation}
\delta = (-122 ^{+41}_{-33})^{\rm o}.
\label{global-fit-delta}
\end{equation}
It is expected that the angle $\alpha$ will be measured with an accuracy goal $\pm 1^{\rm o}$
in ongoing $B$-factories, LHCb and Belle II,
and the CP phase $\delta$ will be measured with the accuracy at the level of 20\% \cite{Fukasawa:2016yue}
(if the phase is near $-90^{\rm o}$)
in near future experiments of long-baseline neutrino oscillations \cite{Abe:2014oxa,Acciarri:2016crz}.
It is true that such angles are just parameters in the standard model and can be nearly right accidentally,
but it is interesting if there is a reason to make them (nearly) right.
In fact, such possibility is studied in the literatures
for the unitarity triangle in the quark mixing matrix \cite{Fritzsch:1995nx,Koide:2004gj,Masina:2006ad,Harrison:2009bb,Xing:2009eg,Antusch:2009hq,Kim:2011yz}
and for the Dirac CP phase in the neutrino mixing \cite{Antusch:2009hq,Kim:2011yz,Aizawa:2005jc,Dicus:2010yu,Wu:2012ria,Zhang:2017sag}.
We expect that a small deviation from a special value (right angle in this case)
may contain a hint of the flavor physics how the flavor mixing is generated and CP is violated.

In this paper, we consider the origin of the nearly right angles of the phases
in flavor models.
The pure imaginary vacuum expectation values (vevs) can be obtained by a singlet scalar potential
with $Z_2$ symmetry.
We examine where the pure imaginary should appear in the Yukawa coupling
in the models, in which the Yukawa matrices are forbidden at the leading order by a flavor symmetry
but can be generated by integrating out heavy vector-like matter fields
(Frogatt-Nielsen(-like) mechanism \cite{Froggatt:1978nt}).
For the purpose, we first study flavor covariant forms of the mass matrices of quarks and leptons.
It is often assumed that the matrices have a texture~\cite{Fritzsch:1977vd,Branco:1994jx} (or nearest neighbor interaction form)
in which several elements are zero as an ansatz. 
However, such texture can be always obtained by unphysical flavor transformation without loss of generality \cite{Branco:1988iq},
and it is unclear why such special flavor frame fixed by the unphysical rotation has a physical meaning
to discuss the origin of the right unitarity triangles.
Therefore, we discuss the origin in the covariant form under the flavor transformation,
to make the results available in many types of flavor models.
We investigate the approximate diagonalization matrices of the fermion mass matrices,
and how the CKM and PMNS mixing matrices can be written.
We construct a flavor model where the pure imaginary vev can generate the nearly right unitarity triangle,
and we examine the possible deviation from the right-angled phases in the quark and lepton sectors.

This paper is organized as follows:
In section 2, we study the general description of the 
quark and lepton mixing matrices 
using the flavor covariant expression of the mass matrices,
and see where the nearly right-angled phase should exist in the parametrization.
In section 3, we study that the pure imaginary coupling
in the flavor covariant form of the mass matrices can generate the nearly right unitarity triangle
in a simple manner,
and examine the deviation of the right angle in a model.
We also study how the nearly right-angled Dirac CP phase in the neutrino mixings
can be obtained in the scenario of the quark-lepton unification.
Section 4 is devoted to the conclusion and discussion.

\section{General description of the mixing matrices}

Using three independent vectors $\eta_a$, any $3\times 3$ Hermitian matrix $A$ can be written as
$A_{ij} = \sum_{a,b} \eta^i_a A_{ab} \eta^{j*}_{b}$.
Diagonalizing the matrix $A_{ab}$, the Hermitian matrix can be expressed
as
\begin{equation}
A_{ij} = \sum_a A_a \xi^i_a \xi^{j*}_{a},
\end{equation}
where $\xi_a$ is a linear combination of the basis vector $\eta_a$.
As a trivial example, if $\eta_a$ is chosen to be the normalized orthogonal basis $\eta_a^i = \delta_{a}^i$,
$A_a$ and $\xi_a$ are the eigenvalues and eigenvectors of the matrix $A$.
%
%
Using the theorem of the linear algebra, 
the matrices of 
$M_f M_f^\dagger$ ($f = u,d,e$), 
where $M_u$, $M_d$ and $M_e$ are up-, down-type quark and charged-lepton mass matrices,
can be parametrized (omitting the flavor indices) as
\begin{equation}
M_f M_f^\dagger = \sum_a \rho_a^f \xi^f_a (\xi^f_a)^*.
\end{equation}
%
Suppose we choose naive bases so that 
the vectors $\xi_a$ are not nearly parallel 
(i.e. $\cos\theta_{ab} = |\xi_a \cdot \xi_b|/|\xi_a||\xi_b|$ is not close to 1),
$\rho_a^{f}$ are hierarchical (relating to the squared fermion masses),
$\rho^{f}_1 \ll  \rho^{f}_2 \ll \rho^{f}_3$.

The (real) vector $\xi_3 = (a,b,c)^T$ is rotated to $(0,0,\sqrt{a^2+b^2+c^2})^T$
by using a unitary matrix $U_0$, 
\begin{eqnarray}
U_0 &=& 
\left(
\begin{array}{ccc}
1 & 0 & 0 \\
0 & \cos\theta_a & -\sin\theta_a \\
0 & \sin\theta_a & \cos\theta_a 
\end{array}
\right)
\left(
\begin{array}{ccc}
\cos\theta_s & -\sin\theta_s & 0 \\
\sin\theta_s & \cos\theta_s & 0 \\
0 & 0 &1
\end{array}
\right) \nonumber
 \\
&=&
\left(
\begin{array}{ccc}
\cos\theta_s & -\sin\theta_s & 0 \\
\cos\theta_a \sin\theta_s & \cos\theta_a \cos\theta_s & -\sin\theta_a \\
\sin\theta_a \sin\theta_s  & \sin\theta_a \cos\theta_a & \cos\theta_a 
\end{array}
\right),
\label{U0}
\end{eqnarray}
where $\tan\theta_s = a/b$ and $\tan\theta_a = \sqrt{a^2+b^2}/c$,
and the direction of the third generation is (nearly) fixed.
The two angles $\theta_s$ and $\theta_a$ are generically large,
which can be the origin of the large solar and atmospheric neutrino mixings.
In the quark sector, on the other hand,
the left-right symmetry or horizontal flavor symmetry
can make the large angles to be cancelled by assuming $\xi_3^u \simeq \xi_3^d$,
and the quark mixing angles are small.
This is the zero-th order explanation of the observed nature of the quark and lepton mixings \cite{Dutta:2005bb},
and it can be applied to the quark-lepton unified models~\cite{Dutta:2004zh}.
The bi-large neutrino mixings can be also explained in the context of SU(3) horizontal flavor symmetry \cite{Kitano:2000xk}.

\subsection{Quark sector}

The quark mass matrices after removing the large mixing angles by $U_0$ rotation can be parametrized
as
\begin{eqnarray}
M_u M_u^\dagger &\propto&
\left( 
 \begin{array}{c}
  0 \\ 0 \\ 1 
  \end{array}
\right)
\left( 
 \begin{array}{ccc}
  0 & 0 & 1 
  \end{array}
\right)
+ \frac{m_c^2}{m_t^2} 
\left( 
 \begin{array}{c}
  x_u \\ y_u \\ z_u 
  \end{array}
\right)
\left( 
 \begin{array}{ccc}
  x_u^* & y_u^* & z_u^* 
  \end{array}
\right)
+  \frac{m_u^2}{m_t^2}  \xi_1^u \xi_1^{u\dagger},
\\
M_d M_d^\dagger &\propto&
\left( 
 \begin{array}{c}
  0 \\ \epsilon \\ 1 
  \end{array}
\right)
\left( 
 \begin{array}{ccc}
  0 & \epsilon^* & 1 
  \end{array}
\right)
+ \frac{m_s^2}{m_b^2} 
\left( 
 \begin{array}{c}
  x_d \\ y_d \\ z_d 
  \end{array}
\right)
\left( 
 \begin{array}{ccc}
  x_d^* & y_d^* & z_d^* 
  \end{array}
\right)
+ \frac{m_d^2}{m_b^2}  \xi_1^d \xi_1^{d\dagger},
\end{eqnarray}
Without loss of generality, one can choose
$(\xi_3^u)^{1,2} = (\xi_3^d){}^{1} = 0$.
We suppose that the last term can be negligible to diagonalize the matrices as an approximation
due to the fermion mass hierarchy.
This approximation is not bad if the basis is chosen appropriately
(i.e. $|\xi_a \cdot \xi_1|/|\xi_a| |\xi_1|$ ($a=2,3$) are not close to 1).
The diagonalization matrices ($V_{f} M_{f} M_{f} V_{f}^\dagger$ is diagonal)
can be expressed
as
\begin{eqnarray}
V_u &=& V_u^{23} (\theta_u^{23}) V_u^{12} (\theta_u), \\
V_d &=& V_d^{23} (\theta_d^{23}) V_d^{12} (\theta_d) V_d^{23} (\theta_\epsilon),
\end{eqnarray}
where $V_d^{23}(\theta_\epsilon)$ rotates the component $\epsilon$ out in $M_d M_d^\dagger$,
and $V_{u,d}^{12} (\theta_{u,d})$ rotate the $x_{u,d}$ component out. 
After the rotations to make the $M_{u,d}M_{u,d}^\dagger$ matrices only in 2-3 blocks,
$V_{u,d}^{23}(\theta_{u,d}^{23})$ diagonalize them.
Choosing the vector $\xi_2^{u,d}$ appropriately to make $y_{u,d} \sim z_{u,d}$ (or  $y_{u,d} \gg z_{u,d}$),
we find that the angles $\theta_{u,d}^{23}$ are small, and we neglect them.
We find
\begin{eqnarray}
V_{\rm CKM} &=& V_u V_d^\dagger \simeq  V_u^{12} (\theta_u) V_d^{23}(\theta_\epsilon)^\dagger V_d^{12} (\theta_d)^\dagger 
\nonumber
\\
&=&
\left(  
 \begin{array}{ccc}
  c_u & - s_u e^{i \alpha_u} & 0\\
  s_u e^{-i\alpha_u} & c_u &0 \\
  0&0 & 1
 \end{array}
\right)
\left(  
 \begin{array}{ccc}
  1 &0 &0 \\
  0& c_\epsilon &  s_\epsilon  \\
  0& -s_\epsilon & c_\epsilon  \\
 \end{array}
\right)
\left(  
 \begin{array}{ccc}
  c_d &  s_d e^{i \alpha_d} &0 \\
  -s_d e^{-i\alpha_d} & c_d  &0 \\
  0& 0& 1
 \end{array}
\label{approx}
\right),
\end{eqnarray}
where $\alpha_{u,d}$ are phases for the respective rotations,
and $c_u =\cos\theta_u$, $s_u = \sin\theta_u$ and so on.
By a phase rotation, $\epsilon$ can be made to be real without loss of generality as it can be seen in the parametrization.
One can find that
$\tan\theta_\epsilon = \epsilon$,
$\tan\theta_u = |x_u/y_u|$,
$\alpha_u = {\rm arg}\, x_u/y_u$, and so on.
We calculate the matrix as
\begin{equation}
V_{\rm CKM} \simeq
\left(  
 \begin{array}{ccc}
  c_u c_d + c_\epsilon s_u s_d e^{i(\alpha_u-\alpha_d)} & c_u s_d e^{i\alpha_d}- c_d c_\epsilon s_u e^{i \alpha_u} & 
  - s_\epsilon s_u e^{i \alpha_u}\\
  -c_u  c_\epsilon s_d e^{-i \alpha_d} + c_d s_u e^{-i\alpha_u} & c_d c_\epsilon c_u + s_d s_u e^{i(\alpha_d-\alpha_u)} &
  c_u s_\epsilon \\
  s_d s_\epsilon e^{-i \alpha_d} & -c_d s_\epsilon & c_\epsilon
 \end{array}
\right),
\end{equation}
\begin{equation}
\epsilon \simeq V_{cb}, \quad \tan\theta_u = |V_{ub}/V_{cb}|, \quad
\tan\theta_d = |V_{td}/V_{ts}|,
\end{equation}
and we obtain the Cabibbo angle as
\begin{equation}
\sin\theta_C \simeq |s_d e^{i\alpha_d} - s_u e^{i\alpha_u}|.
\end{equation}

The sides of the unitarity triangle are given as
\begin{eqnarray}
A &\equiv& V_{cd} V_{cb}^* =  c_u s_\epsilon e^{-i \alpha_d} (-c_\epsilon c_u s_d + c_d s_u e^{i(\alpha_d-\alpha_u)}) ,\\
B &\equiv& V_{ud} V_{ub}^* =  -s_u s_\epsilon e^{-i \alpha_d} (c_\epsilon s_u s_d + c_d c_u e^{i(\alpha_d-\alpha_u)}) ,\\
C &\equiv& V_{td} V_{tb}^* =   s_d s_\epsilon c_\epsilon e^{-i \alpha_d} ,
\end{eqnarray}
satisfying $A+B+C = 0$.
The angles of the triangle are given as
\begin{equation}
\alpha (= \phi_2 ) = {\rm arg}\, (-C/B), 
\quad
\beta (= \phi_1 ) = {\rm arg}\, (-A/C),
\quad
\gamma (= \phi_3 ) = {\rm arg}\, (-B/A),
\end{equation}
We obtain
\begin{equation}
\alpha = -{\rm arg}\, \left(s_u^2 + \frac{s_u c_u c_d}{s_d c_\epsilon}e^{i(\alpha_d -\alpha_u)}\right),
\end{equation}
and one finds that the physical phase in this parametrization is nearly equal to the $\alpha$ angle,
\begin{equation}
\alpha \simeq \alpha_u - \alpha_d
\end{equation}
under the convention where $\theta_{u,d,\epsilon}$ are in the first quadrant.
For $\alpha_u-\alpha_d = \pi/2$,
it is interesting to note that
 the $\alpha$ angle shifts from $\pi/2$ a little bit 
at the order of $s_u s_d \simeq 0.02$ (rad) $= 1^{\rm o}$ to the first quadrant because
${\rm Re}\, (-C/B)$ is positive.


We remark that the approximate expression in Eq.(\ref{approx})
is obtained not by assuming a texture of the mass matrices.
In Ref.\cite{Couture:2009it}, authors discuss
a ``two-angle ansatz" by using an exact formula of the diagonalization unitary matrix of the hermitian matrix, 
$M_f M_f^\dagger$.
They examine if the observed quark masses and mixings can be consistent with the ansatz
under which the hermitian matrices can be diagonalized by two angles.
The condition of the ansatz is independent to the texture-zero flavor frames.
In this paper, we argue that the matrix can be written by the superposition of the 
rank 1 matrices, and thus, the CKM matrix can be written as Eq.(\ref{approx}) approximately
due to the hierarchy of the squared masses.
Surely, the superposition of rank 1 matrices is not unique,
and so, we just claim that there exists a suitable choice of $\xi_i$ 
to express the CKM matrix as given in Eq.(\ref{approx}).

\subsection{Lepton sector}

Since the model dependence in the neutrino mass matrix ${\cal M}_\nu$ is large, let us express 
the charged lepton mass matrix in the basis where ${\cal M}_\nu$ is diagonal:
\begin{eqnarray}
M_e M_e^\dagger &\propto&
\left( 
 \begin{array}{c}
  a \\ b \\ c 
  \end{array}
\right)
\left( 
 \begin{array}{ccc}
  a & b & c 
  \end{array}
\right)
+ \frac{m_\mu^2}{m_\tau^2} 
\left( 
 \begin{array}{c}
  u \\ v \\ w 
  \end{array}
\right)
\left( 
 \begin{array}{ccc}
  u^* & v^* & w^* 
  \end{array}
\right)
+  \frac{m_e^2}{m_\tau^2}  \xi_1^{e\prime} \xi_1^{e\prime\dagger},
\\
{\cal M}_\nu &=& {\rm diag}.\, (m_1, m_2, m_3),
\end{eqnarray}
where $a,b,c$ can be made to be real by pushing their phases to the Majorana neutrino masses, $m_i$.
By rotating the charged lepton matrix by $U_0$ in Eq.(\ref{U0}),
we express
\begin{equation}
U_0 M_e M_e^\dagger U_0^T
\propto
\left( 
 \begin{array}{c}
  0 \\ 0 \\ 1 
  \end{array}
\right)
\left( 
 \begin{array}{ccc}
  0 & 0 & 1 
  \end{array}
\right)
+ \frac{m_\mu^2}{m_\tau^2} 
\left( 
 \begin{array}{c}
  x_e \\ y_e \\ z_e 
  \end{array}
\right)
\left( 
 \begin{array}{ccc}
  x_e^* & y_e^* & z_e^* 
  \end{array}
\right)
+  \frac{m_e^2}{m_\tau^2}  \xi_1^{e} \xi_1^{e\dagger}.
\end{equation}
As in the quark sector, we neglect the last term to diagonalize the matrix as an approximation,
and then, we obtain the neutrino mixing matrix as
\begin{equation}
U_{\rm PMNS} = V_e^{23} (\theta_e^{23}) V_e^{12}(\theta_e) U_0 (\theta_s, \theta_a).
\label{PMNS0}
\end{equation}
Similarly to the quark sector, we neglect the rotation by $\theta_e^{23}$, and we obtain
\begin{eqnarray}
U_{\rm PMNS} &\simeq & \left(  
 \begin{array}{ccc}
  c_e & - s_e e^{i \alpha_e} & 0\\
  s_e e^{-i\alpha_e} & c_e  &0\\
  0&0 & 1
 \end{array}
\right)
\left(  
 \begin{array}{ccc}
  1 & 0&0 \\
  0& c_a &  -s_a  \\
  0& s_a & c_a  \\
 \end{array}
\right)
\left(  
 \begin{array}{ccc}
  c_s &  -s_s &0 \\
  s_s& c_s  &0\\
  0 &0 & 1
 \end{array}
\right), 
\nonumber
\\
&=&
\left(  
 \begin{array}{ccc}
  c_e c_s - c_a s_s s_e e^{i\alpha_e} & - c_e s_s - c_a c_s s_e e^{i \alpha_e} & 
  s_e s_a e^{i \alpha_e}\\
  c_a  s_s c_e  + c_s s_e e^{-i\alpha_e} & c_a c_s c_e - s_e s_s e^{-i\alpha_e} &
  -c_e s_a \\
  s_a s_s  & c_a s_a & c_a
 \end{array}
\right),
\label{PMNS}
\end{eqnarray}
where $\tan\theta_e = |x_e/y_e|$ and $\alpha_e = {\rm arg}\, x_e/y_e$.
Comparing the standard convention,
\begin{equation}
U_{\rm PMNS}
=
\left(
\begin{array}{ccc}
 c_{13}c_{12} & s_{12} c_{13} & s_{13} e^{-i\delta} \\
 -s_{12}c_{23} - c_{12}s_{23}s_{13}e^{i\delta} & c_{12} c_{23} - s_{12} s_{23} s_{13} e^{i\delta} & s_{23} c_{13} \\
 s_{12} s_{23} - c_{12} c_{23}s_{13}e^{i\delta} & -c_{12}s_{23}- s_{12}c_{23} s_{13} e^{i\delta} & c_{23} c_{13}
\end{array}
\right),
\end{equation}
where the mixing angles $\theta_{ij}$ are defined to be in the first quardrant as a convention,
we obtain
\begin{equation}
s_{13} = s_e s_a,
\quad
t_{23} = c_e t_s,
\quad
t_{12} = \left| \frac{c_e s_s + c_a c_s s_e e^{i \alpha_e}}{c_e c_s- c_a s_s s_e e^{i\alpha_e} }\right|.
\end{equation}
To extract the Dirac CP phase $\delta$, 
we use Jarskog invariant which does not change under the unphysical phase rotation,
\begin{equation}
J \equiv U_{e2} U_{\mu 3} U_{e3}^* U_{\mu 2}^*,
\end{equation}
\begin{equation}
J_{CP} = {\rm Im} \, J = c_{13}^2 s_{13} c_{12} s_{12} c_{23} s_{23} \sin\delta.
\end{equation}
We obtain
\begin{equation}
J_{CP} =  -(c_e^3 c_a c_s s_a^2 s_s s_e+ c_a c_e c_s s_a^2 s_s s_e^3)  \sin\alpha_e,
\end{equation}
and in the limit $s_e \to 0$, one can find that $\sin \delta \simeq -\sin \alpha_e$
under the convention where $\theta_{e,s,a}$ are in the first quadrant.
The quadrant of the $\delta$ phase can be checked by using 
\begin{equation}
{\rm Re}\, J + c_{13}^2 s_{13}^2 s_{12}^2 s_{23}^2 = c_{13}^2 s_{13} c_{12} s_{12} c_{23} s_{23} \cos\delta,
\end{equation}
and we can calculate ${\rm Re}\, J$ 
as
\begin{equation}
{\rm Re}\, J = c_e^2 s_e^2 s_a^2 (c_a^2 c_s^2 -s_s^2) + (c_e^3 c_a c_s s_a^2 s_s s_e- c_a c_e c_s s_a^2 s_s s_e^3)\cos\alpha_e.
\label{ReJ}
\end{equation}
It is interesting to note that one obtains ${\rm Re}\, J =0$
 if $\cos\alpha_e=0$ and $U_0$ is tri-bimaximal ($t_a = 1$, $t_s^2 =1/2$).
Supposing $\alpha_e = \pi/2$ and using the experimental data
($\theta_{23} \simeq 45^{\rm o}$, $\theta_{12} \simeq 33.5^{\rm o}$),
we obtain that the $\delta$ phase shifts from $-\pi/2$ depending on $\theta_e$,
and $\delta$ is in the forth quadrant ($\cos\delta>0$).
For $s_{13} = 0.146$ \cite{Capozzi:2018ubv}, we obtain $\delta \simeq -80^{\rm o}$.

\section{Model building of the nearly right-angled CP phases}

In the parametrization given in the previous section,
we find that the choice of $\alpha_u-\alpha_d \simeq \pi/2$ and $\alpha_e \sim \pi/2$
can reproduce the observations naively.
One can expect that a pure imaginary number in the mass matrices can provide such a situation.
However, the expression in the previous section depends on the basis and
it is not clear where we need the pure imaginary in the mass matrix $M_f$ (not in $M_f M_f^\dagger$).
In this section, we describe the mass matrix and see where we need it in the heuristic construction.

\subsection{Nearly right unitarity triangle in quark sector}

In order to illustrate an essential point, we first suppose that the mass matrices are rank 2
and are given by using column vectors $\phi_a$ as follows:
\begin{equation}
M_u  \propto \phi_3 \phi_3^T + \lambda_u \phi_2^u \phi_2^{uT},
\qquad
M_d  \propto \phi_3 \phi_3^T + \lambda_d \phi_2^d \phi_2^{dT}.
\label{toy}
\end{equation}
In general, it is not necessarily that the matrices are symmetric and the transposed vectors can be
different from $\phi_a$. Here we just assume that they are symmetric just for simplicity to describe.
One can find that
\begin{equation}
M_d M_d^\dagger 
\propto \left(
\begin{array}{cc}
 \phi_2^d & \phi_3
 \end{array}
\right)
\left(
 \begin{array}{cc}
  |\lambda_d|^2 \phi_2^{dT} \phi_2^{d*} & \lambda_d \phi_2^{dT} \phi_3^* \\
  \lambda_d^* \phi_3^T \phi_2^{d*} & \phi_3^T \phi_3^*
 \end{array}
\right)
\left(
\begin{array}{c}
\phi_2^{d\dagger}
\\
\phi_3^\dagger
\end{array}
\right),
\end{equation}
where we note $\phi_3^T \phi_3^*$ is just a number.
Diagonalizing the $2\times 2$ matrix in the middle,
one can rewrite it as
\begin{equation}
M_d M_d^\dagger \propto \rho_2^d \xi_2^{d} \xi_2^{d\dagger} + 
\rho_3^d \xi_3^{d} \xi_3^{d\dagger},
\end{equation}
where $\rho_a^d$ $(a=2,3)$ are eigenvalues of the middle $2\times 2$ matrix
 and $\xi_a^{d}$ $(a=2,3)$ are linear combinations of $\phi_3$ and $\phi_2^d$.
Since $\lambda_d$ is small for the fermion mass hierarchy, 
the mixing angle $\epsilon_d$ of the middle matrix is expected to be small,
and we can write as
\begin{equation}
\xi_3^{d} \simeq \phi_3 + \epsilon_d \phi_2^d =
 \left(
 \begin{array}{ccc}
  \epsilon_d x_d \\ \epsilon_d y_d \\ 1+\epsilon_d z_d
 \end{array}
 \right), 
 \quad
 \xi_2^{d} \simeq - \epsilon_d^* \phi_3 + \phi_2^d =
 \left(
 \begin{array}{ccc}
  x_d \\ y_d \\ z_d - \epsilon_d^*
 \end{array}
 \right), 
\end{equation}
where we define $\phi_3=(0,0,1)^T$ and $\phi_2^d = (x_d,y_d,z_d)^T$.
The same expression can be written for the up-type one.
One can easily find that the CKM matrix in this case can be written as
\begin{eqnarray}
V_{\rm CKM} &=&  V_{23}^u (\theta^u) V_{12}^{ud} (\theta^{ud}) V^d_{23}(\theta^d)^{\dagger}
\nonumber
 \\
&=&
\left(  
 \begin{array}{ccc}
  1 & 0 &0 \\
  0 & c^u & - s^u e^{i \alpha^u} \\
  0 & s^u e^{-i\alpha^u} & c^u  
 \end{array}
\right)
\left(  
 \begin{array}{ccc}
  c^{ud} &  s^{ud}  &0 \\
  -s^{ud} & c^{ud} &0 \\
  0& 0& 1
 \end{array}
\right)
\left(  
 \begin{array}{ccc}
  1 & 0 &0 \\
  0 & c^d & s^d e^{i \alpha^d} \\
  0 &- s^d e^{-i\alpha^d} & c^d 
 \end{array}
\right)
\nonumber
 \\
&=&
\left(
 \begin{array}{ccc}
  c^{ud} & c^d s^{ud} & s^{ud} s^d e^{i\alpha^d}\\
  -c^u s^{ud} & c^{ud} c^d c^u + s^d s^u e^{i(\alpha^u-\alpha^d)} & c^{ud} c^u s^d e^{i \alpha^d} - c^d s^u e^{i \alpha^u} \\
  -s^{ud} s^u e^{-i \alpha^u} & - c^u s^d e^{-i \alpha^d} + c^{ud} c^d s^u e^{-i \alpha^u} & c^d c^u + c^{ud} s^d s^u e^{i(\alpha^d-\alpha^u)}
 \end{array}
 \label{KM}
\right).
\end{eqnarray}
To avoid a confusion of the parameterization in the previous,
we use superscripts of $u,d$ for the mixing angles and phases.
By $V_{12}^u$ and $V_{12}^d$ rotations ($V_{12}^{ud} = V_{12}^u V_{12}^{d\dagger}$), 
both $(\xi_3^{u,d})^1$ and $(\xi_2^{u,d})^1$ elements can be made to be zero.
Once one of $(\xi_a^u)^1$ and $(\xi_a^d)^1$ is made to be zero, the phase of the other can be rotated away unphysically, 
and thus, there is no physical phase in $V_{12}^{ud}$.
After the $V_{12}$ rotation, the remained 2-3 blocks of $M_{u,d} M_{u,d}^\dagger$ are diagonalized by $V_{23}^{u,d}$.
Similarly to the previous section, from this expression, we obtain the $\alpha$ angle of the unitarity triangle as
\begin{equation}
\alpha = {\rm arg}\, 
\left((s^u)^2  + \frac{c^d c^u s^u}{c^{ud}s^d} e^{i(\alpha^d-\alpha^u)}\right),
\label{argalpha}
\end{equation}
and thus, we obtain 
$\alpha \simeq \alpha^d- \alpha^u$.
Because of $\alpha \simeq \pi/2$,
$s^d \simeq |V_{ub}/V_{us}|$ and $s^u \simeq |V_{ts}/V_{us}|$,
we obtain a relation
\begin{equation}
|V_{cb}| \simeq \sqrt{|V_{ub}/V_{us}|^2 + |V_{td}/V_{us}|^2}.
\end{equation}
It is interesting to note 
that Eq.(\ref{KM}) is the same as the original parameterization of the CKM matrix
given by Kobayashi-Maskawa \cite{Kobayashi:1973fv}.
Since $V_{tb}$ has a phase in this parameterization, 
the standard convention is more practical to describe the CP violation in the experiments.
But, it is useful to analyze this toy example and its extension.

Assuming that the components of $\phi_1$, $\phi_2^{u,d}$ are all real, one can easily find 
$\alpha_{u,d} \simeq {\rm arg}\, \epsilon_{u,d} = {\rm arg}\, \lambda_{u,d}$.
Therefore, one can obtain the nearly right-angle $\alpha$ if one of $\lambda_u$ and $\lambda_d$ is pure imaginary and
the other is real.
One can build a model to realize such a situation as follows: The Yukawa coupling is written as
\begin{equation}
(\lambda_3^u \phi_3^i \phi_3^j  + \lambda_2^{u} \phi_2^{ui} \phi_2^{uj}) q_i u^c_j H_u
+ 
\lambda_3^d \phi_3^i \phi_3^j  q_i d^c_j H_d + \lambda_2^{d}  \phi_2^{di} \phi_2^{dj} S q_i d_j^c \Delta_d/M_* ,
\end{equation}
where $q_i$ denote left-handed quark doublets, and $u^c_i, d^c_i$ are right-handed up- and down-type quarks,
and $H_u$, $H_d$ and $\Delta_d$ are Higgs doublets, and $S$ is a gauge singlet.
%
Imposing $Z_2$ symmetry, we obtain that the superpotential of the $S$ field is given as
\begin{equation}
W = M_S S^2 + \frac{1}{M_*} S^4,
\end{equation}
and the vev of $S$ can be pure imaginary.
In this example, $\lambda_d$ in the mass matrix in Eq.(\ref{toy}) is pure imaginary.
One can also construct a model to make $\lambda_u$ is pure imaginary similarly.
If we take into account the relation of muon and strange mass (at unification scale) $m_\mu \sim 3 m_s$,
there is an advantage to employ the additional doublet $\Delta_d$.
There may be several ways to construct a model to obtain the pure imaginary in the Yukawa coupling,
for example, one can build a model so that the mixing of $H_d$ and $\Delta_d$ is pure imaginary.

Assuming that $\lambda_d$ is pure imaginary and $y_d \simeq z_d$,
we find that
the phase of $\alpha_d$ shifts from $\pi/2$ by 
$ \pm 2 s^d \simeq \pm 2 |V_{ub}/V_{us}| \simeq \pm 0.04$ (rad) ($\simeq \pm 2^{\rm o}$).
{}From Eq.(\ref{argalpha}), the $\alpha$ angle decreases by $s_u s_d \, (\simeq 1^{\rm o})$.
Therefore, we obtain
\begin{equation}
\alpha \simeq 87^{\rm o} \qquad {\rm or} \qquad \alpha \simeq 91^{\rm o}.
\label{postdict}
\end{equation}
If we choose $\lambda_u$ to be pure imaginary (instead of $\lambda_d$),
one finds $\alpha^u$ is shifted by $\pm 2 s^u \simeq 2 |V_{ts}/V_{us}|$ $(\simeq 4^{\rm o}$)
and thus, $\alpha$ is shifted from the right angle a little more.

Now we construct more realistic model keeping the nearly right angle $\alpha$ above.
We employ two column vectors
$\phi_3 = (0,0,1)^T$ and $\phi_2 = (0,a,b)^T$
and a vector orthogonal to $\phi_2$ and $\phi_3$
$\phi_1 = (1,0,0)^T$, 
and we consider the mass matrices as
\begin{eqnarray}
M_u  &\propto& \lambda_{33}^u \phi_3 \phi_3^T + \lambda_{23}^u
(\phi_2 \phi_3^T + \phi_3 \phi_2^T) + \lambda_{22}^u \phi_2 \phi_2^{T}
+ \lambda_{33}^u \phi_1 \phi_1^{T},
\\
M_d  &\propto& \lambda_{33}^d \phi_3 \phi_3^T + \lambda_{22}^d \phi_2 \phi_2^{T}
+ \lambda_{12}^d (\phi_1 \phi_2^T + \phi_2 \phi_1^T).
\end{eqnarray}
By the $\lambda_{12}$ term, $\phi_2$ and $\phi_3$
are mixed, and it corresponds to $\phi_2^d \simeq \phi_2 + \lambda_{12}^d/\lambda_{22}^d \phi_1$
in the toy example in Eq.(\ref{toy}),
and it can generate the Cabibbo mixing $\theta_C$.
By assuming the absence of (1,1) element of $M_d$,
one obtains the well-know empirical relation $\sin\theta_C \simeq \sqrt{m_d/m_s}$.
The toy example in Eq.(\ref{toy})
can provide a
relation
$m_s/m_b \sim V_{ub}/V_{us}$ (which agrees well with the observations) for $y_d \sim z_d$,
but provides a wrong relation for $m_c/m_t$.
We add $\lambda_{23}^u$ term to fix it.
We note that $\lambda_{22}^u$ term may not be needed
since observations suggest $m_c/m_t \sim |V_{ts}/V_{cb}|^2$,
 depending on the horizontal scale and the running of the top quark mass \cite{Fusaoka:1998vc}.
One can also add the other mixing terms such as $\phi_1\phi_2^T$ in $M_u$,
but in order not to destroy the nearly right angle $\alpha$, one needs to add the other terms to be properly aligned,
and that is why we just add only $\lambda_{11}^u$ term for the up quark mass.
As one can find in the same way as in the previous, 
one can obtain the nearly right angle $\alpha$, Eq.(\ref{postdict}), if only $\lambda_{22}^d$ is pure imaginary.
One can also assume that the $\lambda_{12}^d$ term can be also pure imaginary.
However, to realize the so called Goergi-Jarskog relation $m_s = m_\mu/3$ and $m_d = 3 m_e$,
there is an advantage to choose that the only $\lambda_{22}^d$ term is generated by the
additional Higgs, $\Delta_d$.

The Yukawa interaction to generate the mass matrices can be constructed by
means of Frogatt-Nielsen-like mechanism with
non-Abelian continuous flavor symmetry (\cite{Kitano:2000xk} and references therein, \cite{King:2001uz}), such as
$SU(3)$, $SU(2)$ (where the second and third generation are in a flavor doublet),
and also with non-Abelian discrete symmetry \cite{Antusch:2009hq,Ma:2001dn}, $S_3$, $S_4$, $A_4$, and so on.
The detail description of the flavor symmetry is beyond the scope of this paper.

\subsection{Nearly right Dirac CP phase in the neutrino mixing}

We suppose that the charged-lepton mass matrix is given similarly to the down-type quark as 
\begin{equation}
M_e  \propto \lambda_{33}^e \phi_3 \phi_3^T + \lambda_{22}^e \phi_2 \phi_2^{T}
+ \lambda_{12}^e (\phi_1 \phi_2^T + \phi_2 \phi_1^T).
\end{equation}
In the previous section, we work on the basis where the neutrino mass matrix is diagonal,
and $\phi_3$ can be made to be $(0,0,1)$ by the $U_0$ transformation.
However, to arrange the first component in the $\phi_2$ vector to be zero,
one needs additional rotation.
In that sense, 
even if we choose that $\lambda_{12}^e/\lambda_{22}^e$ 
is pure imaginary,
the $\alpha_e$ phase in Eq.(\ref{PMNS0}) can be arbitrary due to the additional rotation.
Therefore, though the parametrization in Eq.(\ref{PMNS0}) can contain such a rotation in general,
one needs an additional mixing parameter
in order to specfy the source of the right-angled phase to be
$\lambda_{12}^e/\lambda_{22}^e$.
We parameterize the neutrino mixing PMNS matrix as\footnote{
In this expression, we neglect the 2-3 mixing rotation matrix to diagonalize $M_e M_e^\dagger$.
If the factor 3 for the Georgi-Jarskog relation is taken into account and $\lambda_{22}^e$
is 3 times $\lambda_{22}^d$, the 2-3 mixing angle
can be also 3 times of the 2-3 mixing in the down-type quark mass matrix.
Then,
the $\delta$ phase can be shifted further by about $\pm6^{\rm o}$ compared to the values given in the text,
depending on the phase choice.
The near future long-baseline experiments may not provide the accuracy to distinguish the contributions.
}
\begin{equation}
U_{\rm PMNS} \simeq U_{12} (\theta_e,\alpha_e) U_{23} (\theta_a,\alpha_a) U_{13} (\theta_\nu, \alpha_\nu) U_{12} (\theta_s,\alpha_s),
\end{equation}
where $U_{ij}(\theta,\alpha)$ denotes the $i$-$j$ rotation with a phase $\alpha$.
By rephasing, one can find that the only two combinations of the phases are physical,
\begin{equation}
\bar\alpha_e = \alpha_e - \alpha_s,\qquad
\bar\alpha_\nu = \alpha_\nu - \alpha_s - \alpha_a.
\end{equation}
The conventional mixing angles in $U_{\rm PMNS}$ are given as
\begin{equation}
s_{13} \simeq | s_a s_e -  s_\nu e^{i (\bar \alpha_\nu- \bar\alpha_e)} | ,  \quad
t_{23} \simeq t_a, \quad
t_{12} \simeq \frac{|s_s  + c_a c_s s_e e^{i \bar\alpha_e} |}{|c_s - c_a s_s s_e e^{i\bar\alpha_e}|},
\label{compare}
\end{equation}
where we neglect $s_e s_\nu$ quadratic terms in $t_{12,23}$ and $c_e \simeq c_\nu \simeq 1$.

We now work on the basis where $\phi_3=(0,0,1)^T$ and $\phi_2= (0,a,b)^T$.
Suppose that the neutrino mass matrix ${\cal M}_\nu$ is given as follows 
after diagonalizing 2-3 block,
\begin{equation}
U_{23} {\cal M}_\nu U_{23}^T
\propto
\left(
\begin{array}{ccc}
 0 & x & y \\
 x & z & 0 \\
 y & 0 & w
 \end{array}
\right).
\label{Mneu23}
\end{equation}
If $y=0$, the neutrino mass matrix can be diagonalized by two angles,
i.e. $s_\nu =0$ and $\tan2\theta_s = 2x/z$,
and it returns to the case which we have already studied.
Surely, the relative phase between $\lambda_{12}^e$ and $x$ is unphysical,
and the phase of $z \lambda_{12}^e /(x \lambda_{22}^e)$ will be the phase $\bar \alpha_e$ in this case.
If only $\lambda_{22}^e$ is pure imaginary, one can easily obtain $\bar\alpha_e = \pi/2$,
and the
PMNS phase can be nearly $-90^{\rm o}$.
As we have already studied in the previous section, it shifts by the size of $\theta_{13}$
and\footnote{
This type of the prediction is also obtained in Ref.\cite{Antusch:2009hq}
in a different setup, though the numerical number is different since it is before the
precise measurement of $\theta_{13}$ by the reactor neutrino oscillations.
}
\begin{equation}
\delta \simeq -80^{\rm o}.
\end{equation}

Next we consider the case $y\neq0$.
In this case, the description depends on how the neutrino mass matrix
and the phase structure in the matrix is assumed.
We here describe an idealistic choice of the matrix.
Simply supposing that
the neutrino mass matrix can be given by
\begin{equation}
{\cal M}_\nu \propto \phi_2 \phi_2^T + 
 \lambda_{x}^\nu \phi_x \phi_x^T + \lambda_{1x}^\nu (\phi_1 \phi_x^T + \phi_x \phi_1^T)
+ \lambda_{12}^\nu (\phi_1 \phi_2^T + \phi_2 \phi_1^T),
\end{equation}
where $\phi_x$ is an orthogonal vector to both $\phi_2$ and $\phi_1$, i.e. $\phi_x = (0,b,-a)^T$,
one can find that the components $x,y,z$ are given by $\lambda_{1x}^\nu,\lambda_{12}^\nu,\lambda_x^\nu$
respectively.
Suppose that the $\lambda_{12}^\nu$ term is forbidden by a discrete symmetry,
one can obtain the case $y=0$, which we do not touch in detail in this paper.
If only $y$ is pure imaginary, we obtain the phase $\alpha_\nu = \pm\pi/2$.
After rotating 1-3 generation, a small quantity appears at (2,3) elements, and thus $\bar\alpha_\nu$
can be shifted from $\pm\pi/2$ at the level of one degree, which we neglect in the following discussion.
If $z$ is pure imaginary, on the other hand,
 1-3 rotation to Eq.(\ref{Mneu23}) generates a real number in (1,1) component,
 and thus, $\alpha_s$ shifts from $\pi/2$ by a size of $s_\nu$ (unless the phase of $y$ is $\pi/4$).
Therefore, it is best to choose $\lambda_{12}^\nu$ to be pure imaginary
by using the $Z_2$ symmetry to obtain $\bar \alpha_e$ and/or $\bar \alpha_\nu$ to be $\pm\pi/2$. 
If we do not assume the form of neutrino mass matrix, 
in general, 
$\bar\alpha_{e,\nu}$ are deviated from the right angles depending on
$s_\nu$ and $\sqrt{\Delta m_{12}^2/\Delta m_{23}^2}$,
which can modify the following numerical quantity at the level of $\pm 10^{\rm o}$.



Due to the additional parameter $s_\nu$ to the previous, 
the PMNS $\delta$ phase can take any value even if $\bar \alpha_e$ or $\bar \alpha_\nu$ is $\pm \pi/2$.
As one can imagine from Eq.(\ref{compare}),
$s_{13} \simeq | s_a s_e -  s_\nu e^{i (\bar \alpha_\nu- \bar\alpha_e)} |$,
the modification from the right angle depends on $\arctan (s_a s_e/s_\nu)$
in the case of $\bar \alpha_\nu - \bar\alpha_e = \pm \pi/2$.
We enumerate the qualitative description by classifying the choice of the phases:
\begin{enumerate}
\item $\bar \alpha_e = \pm \pi/2$ and $\bar \alpha_\nu = 0,\pi$

This case can be obtained if only $\lambda_{22}^e$ is pure imaginary and ${\cal M}_\nu$ is real.
We find $\delta \sim \pm\pi/2$ for $s_e \gg s_\nu$, and $\delta \sim 0,\pi$ for  $s_e \ll s_\nu$.
For $\delta \sim \pm\pi/2$, the phase is sensitive to $s_\nu$. 
Naively it can shift by an angle $\sim \arcsin(s_\nu/s_{13})$.


\item $\bar \alpha_e = 0,\pi$ and $\bar \alpha_\nu = \pm \pi/2$

This case can be obtained if $\lambda_{12}^\nu$ is pure imaginary, and 
$\lambda_{12}^e/\lambda_{22}^e \lambda_{1x}^\nu$ and $\lambda_{x}^\nu$ are real.
In this case, $\theta_{12}$ for the solar neutrino oscillation is modified from $\theta_a$ by $s_e$.
We find $\delta \sim 0,\pi$ for $s_e \gg s_\nu$, and $\delta \sim \pm\pi/2$ for  $s_e \ll s_\nu$.
Similarly to above, for $\delta \sim \pm\pi/2$, the phase is sensitive to $s_e$,
and it can shift by an angle $\sim \arcsin(s_a s_e/s_{13})$.

\item $\bar \alpha_e = \pm\pi/2$ and $\bar \alpha_\nu = \pm \pi/2$

Since the phases are aligned, $s_e$ and $s_\nu$ are additive (or subtractive) directly to
obtain the experimental measurement $s_{13} \simeq 0.146$.
Unless the cancellation of those two is not large, 
we find that $\delta \sim \pm\pi/2$ is stable compared to the other two cases.

\end{enumerate}

In the case 3, the PMNS phase is not sensitive to $s_e$ and $s_\nu$
and one can obtain $\delta \simeq -90^{\rm o}$.
For the other two cases, the shift from the right angle is the function of the ratio $s_a s_e/s_\nu$.
Let us suppose the value of $s_e$ to be 
$V_{us}/3$, which can be expected from the Georgi-Jarskog relation.
Then, 
the shift angle can be estimated as 
\begin{equation}
\Delta \delta \simeq \pm\arcsin \frac{s_{23} V_{us}/3}{s_{13}} \simeq \pm 20^{\rm o},
\end{equation}
which can be checked by numerical calculations.
Since the experimental measurements somehow incidentally show an empirical relation, $s_{13} \simeq V_{us}/\sqrt{2}$,
the shift angle is $\sim\arcsin(1/3)$.
To obtain the nearly right angle, 
we take the case 2, $\bar \alpha_e = 0,\pi$ and $\bar \alpha_\nu = \pm \pi/2$, under the choice ($s_e < s_\nu$) .
Then we obtain
\begin{equation}
\delta \sim -70^{\rm o} \qquad {\rm or } \qquad -110^{\rm o},
\end{equation}
if the phase is near $-\pi/2$ rather than $\pi/2$, which is suggested by experiments.

\section{Conclusion and Discussion}

The mass matrix can be written as a linear combination of rank-1 matrices.
We have shown that nearly right unitarity triangle of the CKM mixing matrix can be simply obtained
if the rank-1 matrices are all real and a coefficient which roughly gives the second generation fermion mass
is pure imaginary.
This may imply that the physics to generate the CP violation in the quark mixings
is related to the one to generate the quark mass hierarchy.
We can expect that a variety of models on this issue can be constructed 
by means of Frogatt-Nielsen-like mechanism with discrete symmetry.
We propose a possibility that the Yukawa coupling to a Higgs representation 
which can generate the muon and strange quark mass ratio, which is quoted as Georgi-Jarskog relation,
is the origin of the nearly right unitarity triangle.
Under the assumption,
the $\alpha (=\phi_2)$ angle, which is nearly right angle in the unitarity triangle,
shifts from $90^{\rm o}$ by units of $|V_{ub}/V_{us}|$,
and the ``postdiction" of the angle is
\begin{equation}
\alpha = 87^{\rm o} \quad {\rm or} \quad 91^{\rm o}.
\end{equation}
In this paper, we concentrate on the essential points to realize the nearly right unitarity triangle,
without touching the detail of the rigid model construction, such like the selection of the flavon fields.
That is because the realization of the right unitarity triangle does not depend on the detail of model building,
 and the key structure can be applied to many types of the flavor models.
The deviation from the right angle can depend on the detail of the model building (e.g. the deviation
comes from the up-type quark mass matrix instead of down-type one),
and we may need a further investigation.
It is expected that the $\alpha$ angle will be measured with the accuracy $\sim 1^{\rm o}$ 
at the $B$-factory, which may give us a hint of the models.

The Dirac CP phase in the neutrino mixings
can be also nearly right angle by a choice of the pure imaginary coefficient in the lepton mass matrix.
The CP phase $\delta$ is being measured by the experiments of neutrino oscillations,
and the current global fits may suggest $\delta \sim -90^{\rm o}$, Eq.(\ref{global-fit-delta}).
We summarize the predictions of the model with
 commenting on whether we can obtain the relative opposite signs
of the CP phases in the framework of quark-lepton unification.
In the model, 
the phases which can be nearly equal to $\alpha$ and $\delta$ are the relative phases 
of 2nd/3rd and 1st/2nd generations in the quark and lepton sectors, respectively.
The definition of the mixing matrices are defined as $V_u V_d^\dagger$ and $U_e U_\nu^\dagger$
($d_L{}_\alpha = V_{\rm CKM}{}_{\alpha i} d_L{}_i$ and $\nu_\alpha = U_{\rm PMNS}{}_{\alpha i} \nu_i$
where $\alpha$ is for $SU(2)_L$ current basis and $i$ is for mass basis).
If only the pure imaginary in the down-type quark and charged-lepton mass matrices
 are the source of CP violation and $-3$ factor from the Clebsch-Gordan coefficient
is taken,
the sign is flipped three times totally,
and it could give the relative opposite signs of $\alpha$ and $\delta$
in quark and lepton sectors.
To obtain the opposite signs, the sign of the 1st generation needs to be chosen, which means
that there still freedom to flip the sign of $\delta$,
and thus, in general, the opposite signs of $\alpha$ and $\delta$ are not necessarily predicted
in this framework.
In the setup, the 1-3 neutrino mixing angle $\theta_{13}$, which has been measured by reactor neutrino oscillations,
becomes 1/3 of the experimental measurement,
and then a contribution from the neutrino mass matrix, $s_\nu$, is needed.
One can make (1,2) elements of the mass matrices to be also $-3$,
and then, we can obtain the empirically successful relation,
$\sin\theta_{13} \simeq V_{us}/\sqrt2$.
However, then, the electron mass is also three times down quark mass naively,
and one needs a tuning to obtain $m_e \sim 1/3 m_d$.
The tuning can also break the simple realization of the empirical relation $V_{us} \simeq \sqrt{m_d/m_s}$.
If one resolves such tuning, 
the prediction is 
\begin{equation}
|\delta| \simeq 80^{\rm o}.
\end{equation}
%
%
In the quark-lepton unification, 
it seems to be more simply realized that $\theta_{13}$ is rather dominated by $s_\nu$
and the contribution from the charged-lepton is $V_{us}/3\sqrt2$.
In this case, the sign of $\delta$ is not predicted neither since it mainly comes from the neutrino mass matrix,
 but if it is nearly right angle,
there are three distinctive predicition in an idealistic neutrino mass matrix,
\begin{equation}
|\delta| \simeq 70^{\rm o},  \quad {\rm or} \quad 90^{\rm o}, \quad {\rm or} \quad 110^{\rm o}.
\end{equation}
The deviation form the right angle is roughly given by $\arcsin(V_{us}/3\sqrt{2} \sin\theta_{13}) \simeq \arcsin(1/3)$.
If the neutrino mass matrix is loosen from the idealistic choice,
they can be corrected by 10-15\% to them, by the size of $\theta_{13}$ and $\sqrt{\Delta m^2_{12}/\Delta m^2_{23}}$.
The accurate measurements may give us a hint to sort out the solutions.

\section*{Acknowledgement}

This work is supported by grant MOST
106-2112-M-002-015-MY3
of R.O.C. Taiwan.

\end{document}